\documentclass[preprint,12pt, nofootinbib]{revtex4}

\usepackage[brazil, english]{babel}
\usepackage[utf8]{inputenc}
\usepackage{graphicx}
\usepackage{epsfig}
\usepackage{amssymb}
\usepackage{amsmath} 
\usepackage{slashed}
\usepackage[unicode=true,bookmarks=false,breaklinks=false,pdfborder={0 0 1},colorlinks=true]
 {hyperref}
\hypersetup{
 citecolor=blue,linkcolor=blue,urlcolor=blue}

\graphicspath{%
    {converted_graphics/}
    {/}
}

\usepackage{tikz,tikz-3dplot}
\tdplotsetmaincoords{80}{45}

\tikzset{surface/.style={draw=black, fill=white, fill opacity=.6}}

\usepackage{tikz}
\usetikzlibrary{decorations.pathmorphing,patterns}
\begin{document}

\title{Probing the Black Hole Interior with Holographic Entanglement Entropy and the Role of AdS/BCFT Correspondence}
\author{Fabiano F. Santos}
\email[Eletronic address: ]{fabiano.ffs23@gmail.com}
\affiliation{School of Physics, Damghan University, Damghan, 3671641167, Iran.\\
Centro de Ciências Exatas, Naturais e Tecnológicas, UEMASUL, 65901-480, Imperatriz, MA, Brazil.\\
Departamento de Física, Universidade Federal do Maranhão, São Luís, 65080-805, Brazil.}


\begin{abstract}
This work explores the black hole information loss paradox, a fundamental challenge in theoretical physics. It proposes insights using Holographic Entanglement Entropy (HEE) and the AdS/BCFT correspondence within Horndeski gravity. The work revisits the time-dependent behavior of HEE to probe black hole interiors and examines its implications for the Page curve, which describes the entropy evolution of Hawking radiation. It also discusses the relationship between conformal field theory (CFT) microstates and black hole thermodynamics through the AdS/BCFT correspondence, suggesting that only a subset of microstates corresponds to black holes with smooth interiors, while others may involve firewalls. The study extends black hole thermal entropy to time-dependent entanglement entropy, offering a perspective on the interplay between quantum mechanics, thermodynamics, and gravity.
\end{abstract}

\flushbottom
\maketitle

\section{Introduction}

The black hole information loss paradox has remained a cornerstone challenge in theoretical physics since its inception, gaining renewed attention with the introduction of the Page curve. This curve, which tracks the fine-grained entropy of Hawking radiation during black hole evaporation, suggests a potential conflict with the principle of unitarity when the entropy of the radiation exceeds the black hole itself \cite{Hawking:1975vcx, Hawking:1976ra}. This apparent violation of unitarity has spurred the development of numerous theoretical frameworks aimed at resolving the paradox.

\begin{figure}[ht]
\centering
\includegraphics[width=\textwidth]{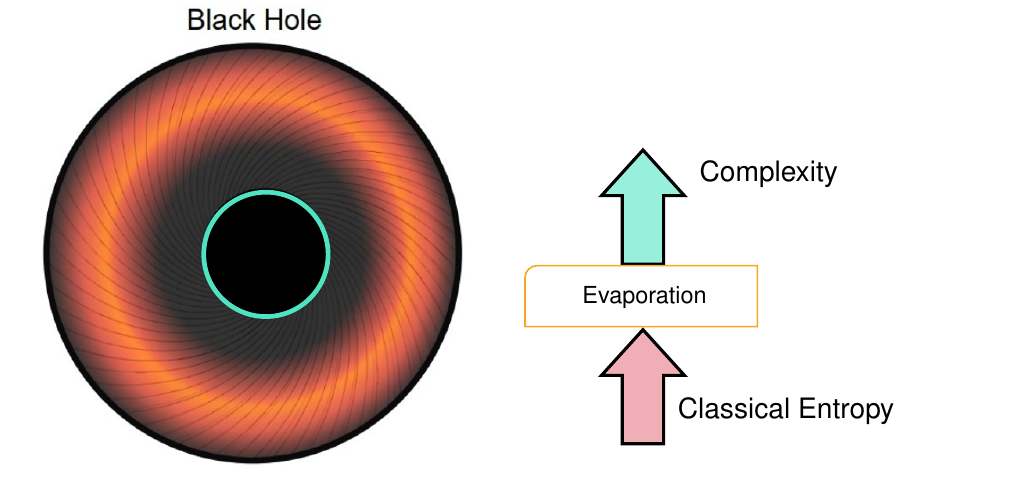}
\caption{Classical entropy has reached its maximum. The interior of a black hole can continue to evolve because of the ever-increasing complexity of its quantum state, meaning there is life after heat death for black holes.}
\label{fig:stream}
\end{figure}

One particularly promising approach is the concept of holographic complexity, which has garnered significant attention in recent years \cite{Susskind:2014rva, Brown:2015bva, Lloyd:2000cry, Brown:2015lvg, Susskind:2018fmx, Brown:2018bms, Brown:2017jil, Brown:2019whu, Brown:2022rwi}. Holographic complexity extends the principles of the AdS/CFT correspondence by proposing that the computational complexity of the boundary field theory encodes critical information about the black hole interior. This framework suggests that black holes continue to emit information even after reaching thermal equilibrium Fig \ref{fig:stream}, offering a potential resolution to the paradox by reconciling the principles of quantum mechanics with the thermodynamic behavior of black holes.

Probing the interior of a black hole presents a significant challenge, as signals from within the event horizon are inaccessible to an external observer. However, in a unitary quantum theory, information must be conserved, raising questions about how this principle reconciles with the apparent inaccessibility of the black hole interior. According to the gauge-gravity duality, all relevant information about the black hole is encoded in the boundary field theory, offering a potential avenue to study the interior indirectly \cite{Santos:2024cvx,Hartman:2013qma}.

\begin{figure}[ht]
\centering
 \includegraphics[width=\textwidth]{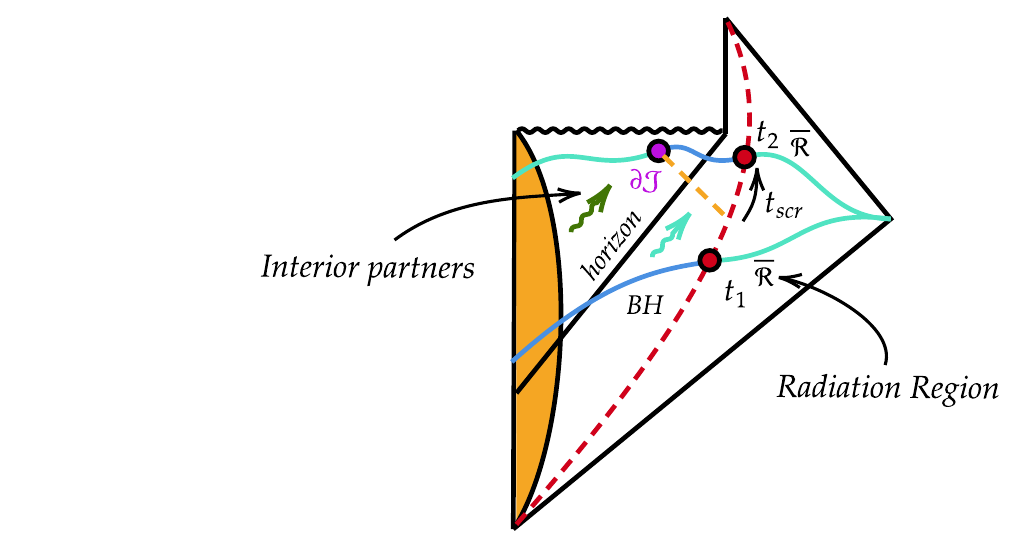}
\caption{The construction of the Double Holographic Model shows that the island is connected to the radiation region through the extra dimension in which the total surface is not disconnected \cite{Santos:2024cvx,Hartman:2013qma}. In the framework of Quantum Extremal Islands, the Holographic Entanglement Entropy of the radiation region $\bar{\mathcal{R}}$. However, field theory with a holographic dual govern entanglement island ${\color{violet}\partial\mathcal{J}}$. This is a manifestation of the $ER=EPR$ idea \cite{Susskind:2016jjb,Susskind:2014yaa}.}
\label{fig:stream1}
\end{figure}

This work presents a summary of specific observables whose gravitational computations involve the black hole interior and Hawking radiation region connected by an island (${\color{violet}\partial\mathcal{J}}$) Fig \ref{fig:stream1}. By "interior," we refer to the region beyond the event horizon, extending toward the singularity, as opposed to the second exterior region of an eternal black hole \cite{Engelhardt:2014gca,Almheiri:2019psy}. The eternal black hole is dual to a thermal state in the conformal field theory (CFT), and a particular black hole microstate formed through gravitational collapse, is characterized by a single asymptotic region.

\begin{figure}[ht]
\centering
\includegraphics[width=\textwidth]{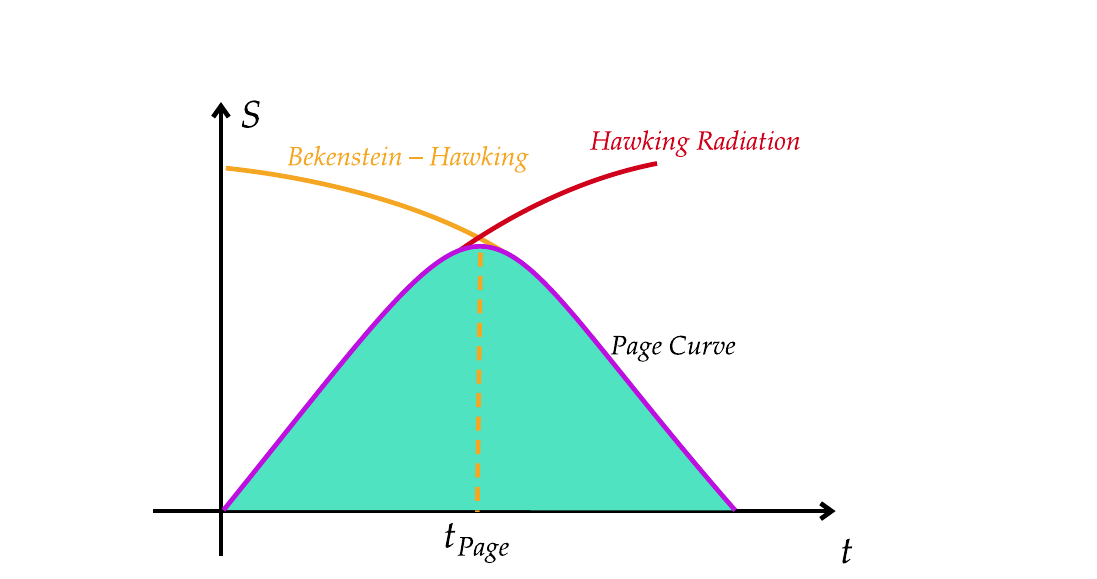}
\caption{The entropy rises due to the increasing entanglement between Hawking quanta. After the Page time, it decreases again and follows the coarse-grained black hole entropy instead.}
\label{fig:stream2}
\end{figure}

The emergence of the Page curve Fig \ref{fig:stream2} from the island phenomenon can be understood by analyzing the entropy dynamics of a black hole in distinct time regimes \cite{2840490,Almheiri:2020cfm}. Eventually, the black hole fully evaporates, and all the information is released. The entropy rises as more entangled particles escape from the black hole. After the Page time, the entropy begins to decline, eventually reaching zero when the black hole has fully evaporated and all the information is released. So, deriving the Page curve essentially resolves the information paradox.

A central tool in the study of entanglement entropy is to compute holographically the area of an extremal surface in anti-de Sitter (AdS) space that terminates on the boundary of the region of interest \cite{Geng:2021mic,Geng:2023qwm,Geng:2021hlu,Geng:2023zhq,Geng:2022dua}. In static spacetimes, extremal surfaces do not penetrate the event horizon \cite{Ryu:2006bv}. However, in time-dependent scenarios \cite{Santos:2024cvx,Hartman:2013qma,Geng:2022dua}, these surfaces can extend into the black hole interior, providing a means to probe its structure. Previous studies have explored this phenomenon using the Vaidya spacetime, which models a collapsing shell of null dust forming a black hole \cite{Hubeny:2007xt,Hubeny:2012ry,Abajo-Arrastia:2010ajo}.

To introduce time dependence, we need to consider an eternal black hole. While it is static under conventional time evolution—forward on one side of the Penrose diagram and backward on the other—we instead evolve the system forward in both exterior regions, creating a time-dependent configuration. This approach serves as a simplified model for thermalization in the strongly coupled dual CFT \cite{Santos:2024cvx,Hartman:2013qma,Geng:2022dua}. Although the two-sided setup may appear artificial, it is relevant to realistic thermalization processes. Certain black hole microstates resemble the eternal black hole outside the horizon but lack the second asymptotic region, corresponding to time-dependent pure states in the CFT undergoing thermalization.

The entanglement entropy of a quantum system is defined by partitioning the system into two subsystems, $\mathcal{A}$ and $\mathcal{B}$, on a fixed-time slice \cite{Santos:2024cvx,Geng:2021mic,Geng:2023qwm,Geng:2021hlu,Geng:2023zhq,Geng:2022dua}. The reduced density matrix of subsystem $\mathcal{A}$ ($L$ is the size of the region $\mathcal{A}$) is obtained by tracing out the degrees of freedom in $\mathcal{B}$,
\begin{eqnarray}
\rho_{\mathcal{A}}=Tr_{\mathcal{B}}\rho_{Total}, 
\end{eqnarray}
and the entanglement entropy is given by 
\begin{eqnarray}
S_{\mathcal{A}}=-Tr(\rho_{\mathcal{A}}log(\rho_{\mathcal{A}})). 
\end{eqnarray}
For a spatial region $\mathcal{A}$ in a two-dimensional CFT, the entanglement entropy in both the ground state and thermal states has been computed using the replica method \cite{Basu:2023jtf}. Time-dependent entanglement entropy has also been studied in the context of global quantum quenches, where the system is prepared in the ground state of a gapped Hamiltonian and then evolved under a new Hamiltonian \cite{Hartman:2013qma}. These studies reveal that entanglement entropy grows linearly with time before saturating at a thermal value, reflecting the thermalization of the subsystem.

On the gravitational side, the entanglement entropy is related to the area of an extremal surface in AdS space that terminates on the boundary of the region $\mathcal{A}$ \cite{Ryu:2006bv}. For static spacetimes, this relationship was proposed by Ryu and Takayanagi and later extended to time-dependent backgrounds. The entanglement entropy is given by the area ($\mathcal{A}(\bar{\gamma})$):
\begin{eqnarray}
S=\frac{\mathcal{A}(\bar{\gamma})}{4G_{N}}
\end{eqnarray}
$\bar{\gamma}$ is the extremal surface and $G_{N}$ is Newton's constant. If multiple extremal surfaces exist, the one with the minimal area is chosen. For large regions ($L>>\beta$, where $\beta$ is the inverse Hawking temperature), the minimal extremal surface initially spans the black hole, connecting the two asymptotic regions. As time evolves, the surface progresses into the interior along specific spacelike slices, leading to linear growth in entanglement entropy, $S_{\mathcal{A}}\alpha\,sv t$, where $s$ is the thermal entropy density and $v$ is the speed of entanglement growth \cite{Hartman:2013qma}.

\section{Methodological route and achievements}

Motivated by the recent application of the AdS/BCFT duality in Horndeski gravity, together with the emergence of AdS/BCFT, and taking into account the significant role of (2+1)$-$dimensional black holes \cite{Santos:2024cvx,Santos:2021orr}. In this work, we establish a description of the time-dependent entanglement entropy to probe the black hole interior using the AdS/BCFT correspondence in Horndeski gravity. Here we present  a  summary  of  the  main  results  achieved  in  this  work:

\begin{itemize}
\item  First, we study the influence of the Horndeski parameters on the AdS/BCFT black hole. Although this result has been present \cite{Santos:2024cvx}, at present, the minimal entropy corresponds to black holes with a smooth interior for which the AdS/BCFT construction is correct. This connection and interpretation it wasn't presented and discussed in \cite{Santos:2021orr}, where we only investigate the thermodynamic properties of the black hole.
\item Second, the \cite{Santos:2024cvx} discusses that  Horndeski parameters significantly alter the behavior of the Page curve compared to standard general relativity, establishing a novel connection between holography and the structure of viable gravitational theories. But here we draw attention to the black hole interior on the same basis of investigation pointed out by \cite{Hartman:2013qma}.
\item Third, in the new discussion presented here, we derive again the results presented by \cite{Santos:2024cvx}, as the entanglement entropy, and we interpret these results to probe the black hole interior, offering new insights into the resolution of the information loss paradox.
\end{itemize}

For the second and third points, using the HEE offers a new perspective on the information loss paradox. Through the AdS/BCFT correspondence, we study time-dependent entanglement entropy and its implications for black hole thermodynamics. The fundamental mechanism involves the interplay between entanglement islands, the Page curve, and the Horndeski gravity framework, which modifies the entropy dynamics.

The summary of this work is: in Sec \ref{BHT} we present the AdS/BCFT black hole \cite{Takayanagi:2011zk,Santos:2024cwf} where we present the minimal entropy for the Horndeski gravity \cite{Santos:2021orr,Santos:2023flb,Santos:2023mee,Santos:2024zoh} and we argue that this minimal entropy could represent a small fraction of CFT microstates. In Sec \ref{BHT1}, we give some arguments where this minimal information could be a candidate to impulse the entropy is now dominated by the area term, $(S_{HM}(t) \approx \frac{\mathcal{A}_h(t)}{4G_N}$) ($(S_{HM}(t)$ is the time depend Hartman-Maldacena entanglement entropy) in a competition between $(S_{HM}(t)$ (for $t\gtrsim \beta$, the entanglement entropy grows linearly in time) and $(S_{Island}(t)$ modify by the Horndeski Lagrangian. In Sec \ref{BHT2}, we present the conclusion and discussions, and future perspectives on this subject. 

\section{The AdS/BCFT black hole}\label{BHT}

Stephen Hawking's work on the black hole information loss paradox has been pivotal in shaping our understanding of quantum gravity and black hole physics \cite{Hawking:1975vcx,Hawking:1976ra}. Initially, Hawking proposed that black holes emit radiation, now known as Hawking radiation, which leads to the gradual evaporation of the black hole. This process appeared to result in the loss of information, violating the principle of unitarity in quantum mechanics. However, in his later work, Hawking suggested a potential resolution to this paradox by introducing the concept of "supertranslations" at the event horizon \cite{Haco:2018ske,Haco:2019ggi,Haco:2019paj}. He argued that these supertranslations, caused by ingoing particles, encode information about the black hole's internal state, allowing it to be preserved in principle, even if practically inaccessible \cite{Hawking:2015qqa}.

\begin{figure}[ht]
\centering
\includegraphics[width=\textwidth]{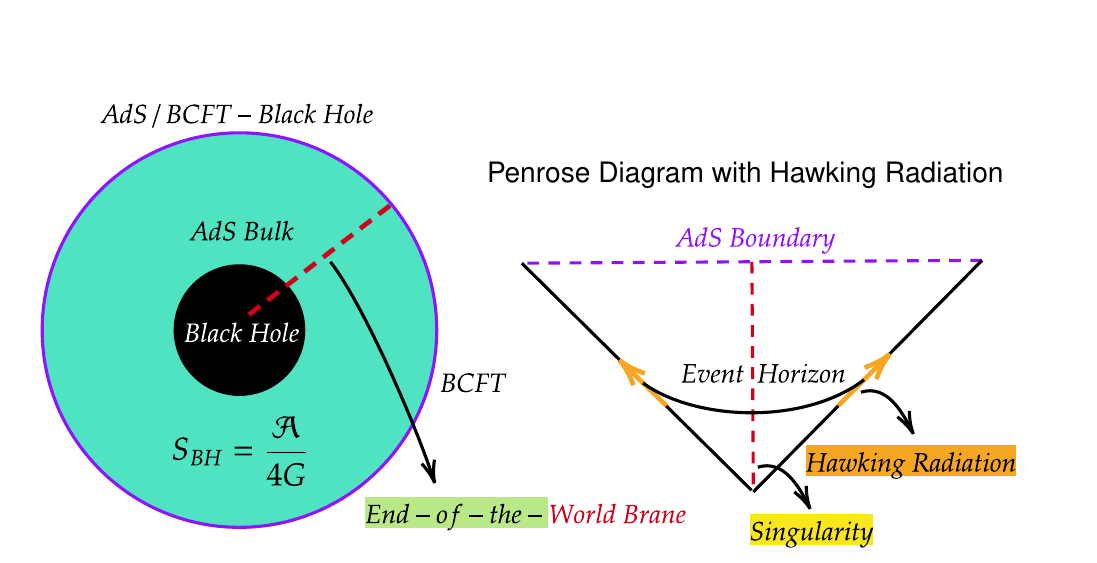}
\caption{{\sl In the left side}, we have the AdS bulk represented as a circular region. The black hole is located in the center of the bulk. The BCFT resides on the boundary of the AdS space. {\sl In the right side}, we have the Penrose diagram with Hawking Radiation: the Penrose diagram shows the causal structure of the black hole. The event horizon, singularity, and AdS boundary are marked. Hawking radiation is represented as arrows propagating outward from the event horizon.}
\label{fig:stream3}
\end{figure}

Further developments in the field have built upon Hawking's ideas. For instance, recent studies have demonstrated that black holes possess "quantum hair," meaning that the quantum state of the external graviton field depends on the internal state of the black hole \cite{Haco:2018ske,Haco:2019ggi,Haco:2019paj}. This implies that Hawking radiation amplitudes are influenced by the internal structure of the black hole, leading to a pure final radiation state that preserves unitarity. This approach challenges the factorization assumption central to the original paradox and suggests that black hole information is encoded in entangled macroscopic superposition states of the radiation \cite{Calmet:2022swf}. These insights have significantly advanced our understanding of the information loss paradox, providing a framework for reconciling the principles of quantum mechanics with the thermodynamic behavior of black holes. With this, we can describe the quantum behavior of the General Relativity (GR) through the AdS/CFT correspondence \cite{Maldacena:1997re,Witten:1998qj}; your extension known as AdS/BCFT correspondence \cite{Takayanagi:2011zk,Santos:2024cwf} opens an avenue with a new scenario to describe the thermodynamics of the black hole with boundary Fig \ref{fig:stream3}.

The framework of the standard General Relativity (GR) \cite{2,3}, provide by Einstein open an avenue in the study of bodies with large mass, and a connection with thermodynamics was performed by Hawking \cite{Hawking:1975vcx,Hawking:1976ra,Haco:2018ske,Haco:2019ggi,Haco:2019paj,Hawking:2015qqa} when computed the entropy of the black hole
\begin{eqnarray}
S=\frac{\mathcal{A}}{4G_{N}},
\end{eqnarray}
but when the temperature tends to zero ($T\to\,0$) all information encoded by the entropy ($S\to\,0$) disappear if the black evaporate completely \cite{4,5}. To address this question that disagrees with the quantum mechanics \cite{Susskind:2016jjb,Susskind:2014yaa}. This question was open for some decades and to resolve it alternatives has been proposed \cite{Susskind:2014rva, Brown:2015bva, Lloyd:2000cry, Brown:2015lvg, Susskind:2018fmx, Brown:2018bms, Brown:2017jil, Brown:2019whu, Brown:2022rwi}, these models show that the black hole emit information after reaching the thermal equilibrium \cite{Santos:2024cvx,Santos:2021orr,Santos:2023flb,Santos:2023mee,Santos:2024zoh}. Thus, the entropies that are treated in General Relativity (GR) are classical. This suggests that the GR must be extend to incorporate or guarantee that some microstates could be describe in some setup. A new setup using an extension of GR, like Horndeski gravity, equipped with AdS/BCFT correspondence, offers a powerful tool to investigate and provide a new entropy to the black hole. 

\begin{figure}[ht]
\centering
\includegraphics[width=\textwidth]{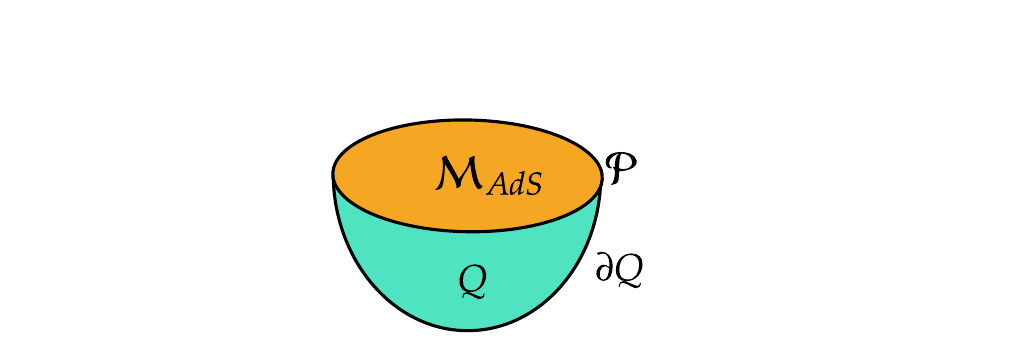}
\caption{AdS/CFT correspondence with boundary hypersurface.}
\label{fig:stream4}
\end{figure}
The AdS/BCFT correspondence, introduced by Takayanagi \cite{Takayanagi:2011zk}, extends the foundational AdS/CFT correspondence \cite{Maldacena:1997re} by incorporating a boundary within the asymptotically AdS spacetime. In this framework, the $d+1$-dimensional AdS space, denoted as $\mathcal{M}_{AdS}$, is augmented by a $d$-dimensional boundary manifold $\partial\mathcal{Q}$, such that the total space is defined as $\mathcal{Q} = \mathcal{M}_{AdS} \cup \partial\mathcal{Q}$ \cite{Santos:2024cvx}. The boundary $\partial\mathcal{Q}$ satisfies the condition $\partial\mathcal{Q} \cap \mathcal{M}_{AdS} = \mathcal{P}$, where $\mathcal{P}$ represents the intersection of the bulk and boundary (see Fig. \ref{fig:stream4}). This extension has attracted significant interest due to its ability to probe a variety of physical phenomena. Entanglement entropy plays a central role in understanding the AdS/BCFT correspondence, as it provides a direct link between quantum information and the geometry of spacetime. In this framework, the entanglement entropy of a subsystem on the boundary theory is holographically related to the area of a minimal surface in the bulk spacetime \cite{Santos:2024cvx}. First, we present the implications of AdS/BCFT correspondence construction with Horndeski gravity for the black hole thermodynamics in three dimensions 
\begin{eqnarray}
ds^{2}_{AdS_3}=\frac{1}{r^{2}}\left(-f(r)dt^{2}+dy^{2}+\frac{dr^{2}}{f(r)}
\right)
\end{eqnarray}
with the Horndeski Lagrangians
\begin{eqnarray}
&&\mathcal{L}^{\mathcal{Q}}_{Horndeski}=R-2\Lambda-\frac{1}{2}(\alpha g_{\mu\nu}-\gamma\,  G_{\mu\nu})\nabla^{\mu}\phi\nabla^{\nu}\phi,\\
&&\mathcal{L}^{\partial\mathcal{Q}}_{Boundary}=K-\Sigma-\frac{\gamma}{4}\big[\nabla_{\mu}\phi\nabla_{\nu}\phi\, (n^{\mu}n^{\nu}+K^{\mu\nu}) -(\nabla \phi)^2)K \big],
\\ 
&&\mathcal{L}^{\mathcal{P}}_{counter\,\,terms}=c_{0}+c_{1}R+c_{2}R^{ij}R_{ij}+c_{3}R^{2}+b_{1}(\partial_{i}\phi\partial^{i}\phi)^{2}+...\,. \nonumber\\
\end{eqnarray}
The entropy of the black hole is given by
\begin{eqnarray}
&&S_{BH}=\frac{\mathcal{A}}{4G_{N}}=\frac{\mathcal{A}_{bulk}+\mathcal{A}_{boundary}}{4G_{N}},\\
&&\mathcal{A}_{bulk}=\pi\,LT\left(1-\frac{\xi}{4}\right),\\
&&\mathcal{A}_{boundary}=4\pi\,LT\left(1-\frac{\xi}{8}\right)+2\pi^2L^3T^2\xi\left(1-\frac{\xi}{8}\right)-7\pi^2L^3\xi\,T^2-\frac{1}{6}\xi\,L,\\
&&\xi=-\frac{1
}{2}\frac{\alpha+\gamma\Lambda}{\alpha}.
\end{eqnarray}
In the limit \( T \to 0 \) with the unitarity parameter \( L \to 1 \), the black hole entropy approaches its minimal value, \( S_{\text{BH}} = -\xi/6 \). This minimal entropy reflects the residual information content of the black hole, even at zero temperature, and is consistent with the holographic principle \cite{Santos:2021orr}. As we pointed out in \cite{Santos:2021orr}, a thermodynamic description of the hole was carried out, and here we will draw attention to the interior of the black hole and its relationship with the residual entropy. Thus, this minimal information shows that a small subset of conformal field theory (CFT) microstates corresponds to black holes with smooth interiors, where the construction remains valid. These microstates are associated with black holes that exhibit a well-defined interior geometry, consistent with the semi-classical description of spacetime. Besides, the majority of microstates correspond to configurations with firewalls, where the construction breaks down due to the phenomenon of state dependence \cite{Basu:2023jtf,Hartman:2013qma,Ryu:2006bv}. This state dependence implies that the mapping between the black hole interior and the boundary CFT is not universal but instead depends on the specific microstate under consideration. Such a framework provides a resolution to several paradoxes in the AdS/CFT correspondence, particularly those related to the information loss paradox and the nature of the black hole interior. By allowing the interior geometry to depend on the state, the theory accommodates both smooth interiors and firewall-like configurations, depending on the microstate.

The state-dependent nature of the mapping between the black hole interior and the boundary CFT also provides a resolution to the firewall paradox. For microstates corresponding to smooth black hole interiors, the entanglement entropy evolves in a manner consistent with semi-classical expectations \cite{Santos:2024cvx}. However, for microstates associated with firewalls \cite{Brown:2017jil}, the entanglement entropy evolution deviates from this behavior, reflecting the breakdown of the semi-classical description. This duality between smooth interiors and firewalls highlights the richness of the CFT microstate structure and its implications for black hole physics.

Furthermore, the AdS/BCFT framework, combined with the thermodynamics of BTZ black holes in Horndeski gravity, provides a powerful tool to study the time evolution of entanglement entropy in more general settings \cite{Santos:2024cvx}. The Horndeski gravity modifies the black hole solutions and their thermodynamic properties, leading to new insights into the relationship between bulk geometry and boundary dynamics. For example, the scalar field from Horndeski gravity can introduce additional contributions to the entanglement entropy, reflecting the bulk scalar dynamics on the boundary CFT. So, the state-dependent mapping between the black hole interior and the boundary CFT, combined with the insights from the AdS/BCFT correspondence and Horndeski gravity, provides a coherent framework to address several paradoxes in black hole physics. The linear growth of entanglement entropy, tied to the growth of spacelike slices in the black hole interior, offers a holographic explanation for the dynamics of information in black holes.


\section{Entanglement competition}\label{BHT1}
The study of entanglement islands in the context of the Karch-Randall braneworld has revealed significant insights into the interplay between quantum gravity and holography \cite{Santos:2024cvx,Geng:2021mic,Geng:2023qwm,Geng:2021hlu,Geng:2023zhq,Geng:2022dua}. It has been generally conjectured and demonstrated that the presence of entanglement islands is incompatible with long-range (massless) gravity. This conclusion aligns with the observation that only a subset of conformal field theory (CFT) microstates correspond to black holes with smooth interiors, where the entanglement island construction remains valid. These findings highlight the nuanced relationship between holographic principles, black hole microstates, and the constraints imposed by gravitational theories.

To explore time-dependent entanglement entropy we consider the AdS$_4$/BCFT$_3$ setup within a three-dimensional Karch-Randall (KR) brane \cite{Karch:2000ct,Randall:1999vf,Brito:2018pwe,Santos:2023eqp} embedded within a four-dimensional black string geometry. By analyzing this setup in the framework of Horndeski gravity, we aim to investigate the influence of modified gravity on the system's behavior, particularly in relation to thermal states and the degrees of freedom associated with the black hole background. The four-dimensional black string is described by the following metric:

\begin{eqnarray}
ds^{2}_{AdS_4} = \frac{1}{r^{2}\sin^{2}(u)}\left(-f(r)dt^{2} + dy^{2} + r^{2}du^{2} + \frac{dr^{2}}{f(r)}\right), \label{me}
\end{eqnarray}
where \( u \in [-\infty, \infty] \), and the asymptotic boundary is located at \( u = -\infty \cup \infty \). The KR brane is embedded at a constant \( u = u_b \) slice \cite{Basu:2023jtf}. 
\begin{figure}[!ht]
\begin{center}
\includegraphics[width=\textwidth]{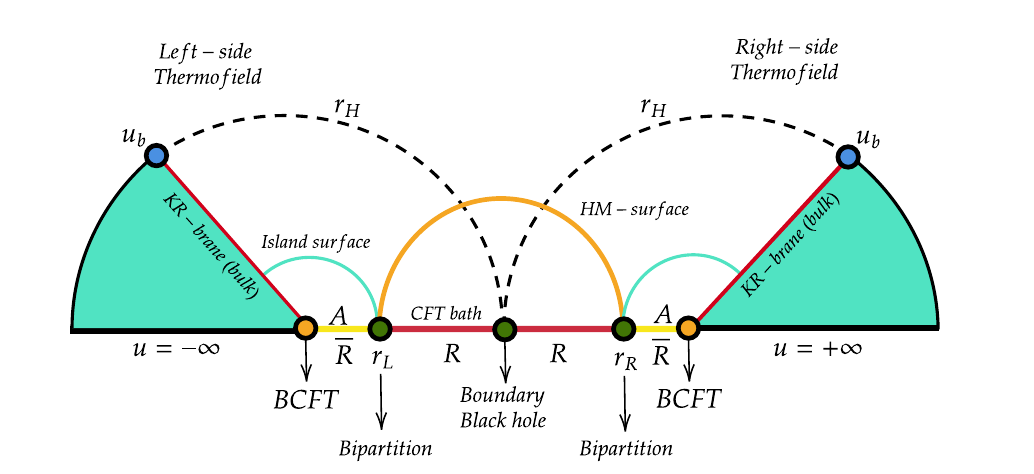}
\end{center}
\caption{Schematic representation of the black string. The $KR$ brane is 
the point $u=u_b$ (red line), the dashed arc at the $r=r_{H}$ represents the black 
string horizon and the orange arc is a Hatman-Maldacena (HM) surface 
\cite{Hartman:2013qma}. Finally, the island surfaces correspond to the blue 
curves.}\label{fig:stream5}
\end{figure}

Figure \ref{fig:stream5} provides a schematic representation of the black string, where the accessible bulk region extends from \( u = u_b \) to \( u = \infty \). Each constant-\( u \) slice of the black string corresponds to an eternal AdS$_4$ black hole with well-defined asymptotic limits. From the perspective of the AdS$_4$/BCFT$_3$ correspondence, the dual field theory is a BCFT$_3$ defined in an AdS$_3$ black hole background. The conformal boundary conditions are imposed at \( r = 0 \), and the metric of the AdS$_3$ black hole is given by:

\begin{eqnarray}
ds^{2}_{AdS_3} = \frac{1}{r^{2}}\left(-f(r)dt^{2} + dy^{2} + \frac{dr^{2}}{f(r)}\right). \label{Q1}
\end{eqnarray}
This construction provides a novel framework to study the interplay between modified gravity and holography, offering insights into the thermal and quantum properties of black hole spacetimes. Considering Fig \ref{fig:stream1} and Fig \ref{fig:stream5}, the Holographic Entanglement Entropy (HEE) for the island connected to the radiation region is the minimum of $S_{HM}$ and $S_{island}$, i.e.  
$S=min(S_{HM},S_{island})$:
\begin{eqnarray}
&&S(\bar{\mathcal{R}})=min(S_{HM},S_{island})=\frac{\mathcal{A}_{island}}{4G}+\frac{\mathcal{A}_{HM}}{4G},\\
&&\mathcal{A}_{island} =2\chi\int^{\rho }_{\rho_{*}}{d\rho}=2\chi( \rho -\rho _{*}),\\
&&\mathcal{A}_{HM}=6\chi\,log\,\left[\frac{2r_h\Delta_r}{r^2}\left(1+\omega(\rho_{
\epsilon})\sqrt{\Delta_L\,\Delta_R}\,\cosh\left(\frac{4\pi\,t}{\beta 
}\right)\right)\right]+2\chi\rho_{\epsilon},\\
&&\chi=\frac{2-\beta_0(1-\beta_0)}{2(1-\beta_0)};\,\,\,\omega=1+\left(\frac{
\alpha-3\gamma}{\gamma}\right)\cosh^2(\rho),\,\,\,(\rho=\rho_{\epsilon}),
\end{eqnarray}
where $\beta_0=\alpha/\gamma\Lambda$, and the parameters are defined in the range $-\infty<\beta_0\leq-1$  with $\alpha,\gamma<0$, or $-1\leq\beta_0<0$  with $\alpha,\gamma>0$ (for more details of this computation see for example \cite{Santos:2024cvx}). The pair of minimal island surfaces that cross from the bipartitions $r_L$ and $r_R$ that correspond to physical branes location at $\rho=\rho_{*}$. With $S_{HM}$ and $S_{island}$, we can derive the Page curve behavior where Page time is obtained as $S_{HM}(t_{Page})=S_{island}$ and Page angle is given by the condition $S_{HM}(t=0)=S_{island}$, respectively:
\begin{eqnarray}
&&t_{Page}=\frac{\beta }{4\pi } cosh^{-1}\left[\frac{e^{-2\rho _{*}} r_{L} r_{R} 
-r_{h}( \Delta _{L} +\Delta _{R})}{2\omega ( \rho _{*}) r_{h}\sqrt{\Delta 
_{L}\Delta _{R}}}\right],\\
&&\rho _{Page} =\frac{1}{2} log\left[\frac{r_{L} r_{R}}{r_{h}\left(\Delta _{L} 
+\Delta _{R} +2\left[ 1+\left(\frac{\alpha -3\gamma }{\gamma 
}\right)\right]\sqrt{\Delta _{L} \Delta _{R}}\right)}\right],
\end{eqnarray}
where $\Delta_L=r_h-r_L,\,\Delta_R=r_h-r_R$.
\begin{figure}[!]
\centering
\includegraphics[scale=0.12]{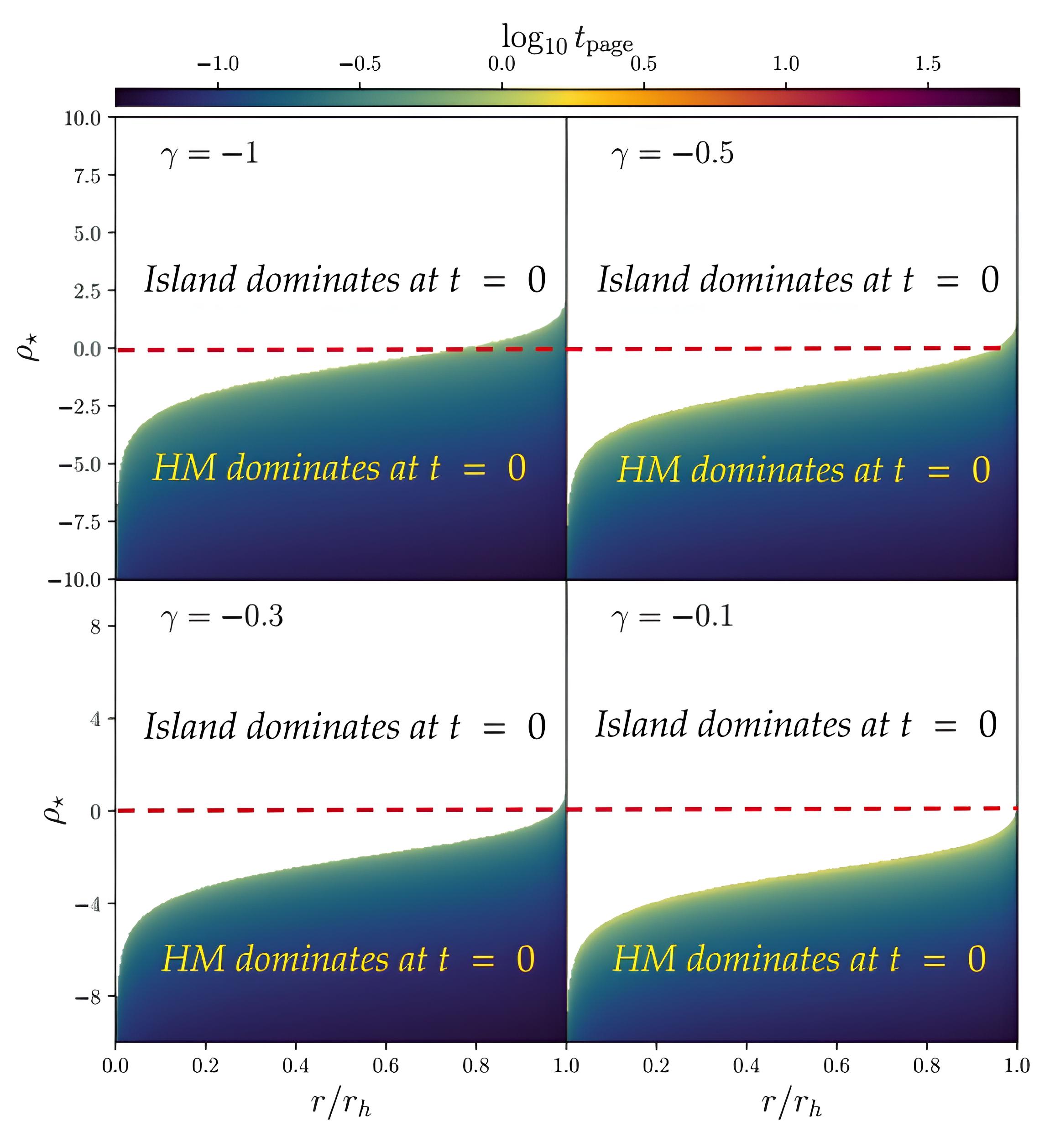}
\caption{The Page time $t_{Page}$ (Extracted and adapted from \cite{Santos:2024cvx}) for various values of $\gamma$ and fixed $\alpha=-8/3$. The points
where the Page angle gives the tensionless brane $\rho=0$ are those where the parameter $\gamma$ is furthest from zero. The fact that $\gamma$ is far 
from its null value shows that for $\rho_{Page}>0$ the island dominates at $t=0$, while for $\rho_{Page}<0$ the HM surface dominates at $t=0$. The
regions showing dominant competition between the island and HM surfaces initially dominate the entropy calculation due to fixing $\alpha$ 
and varying $\gamma$, and reveal the strength of the scalar field contribution. These graphs are equivalent to the density plot along a diagonal slice.} \label{fig:t_page_gamma}
\end{figure}

\begin{figure}[!]
\centering
\includegraphics[scale=0.12]{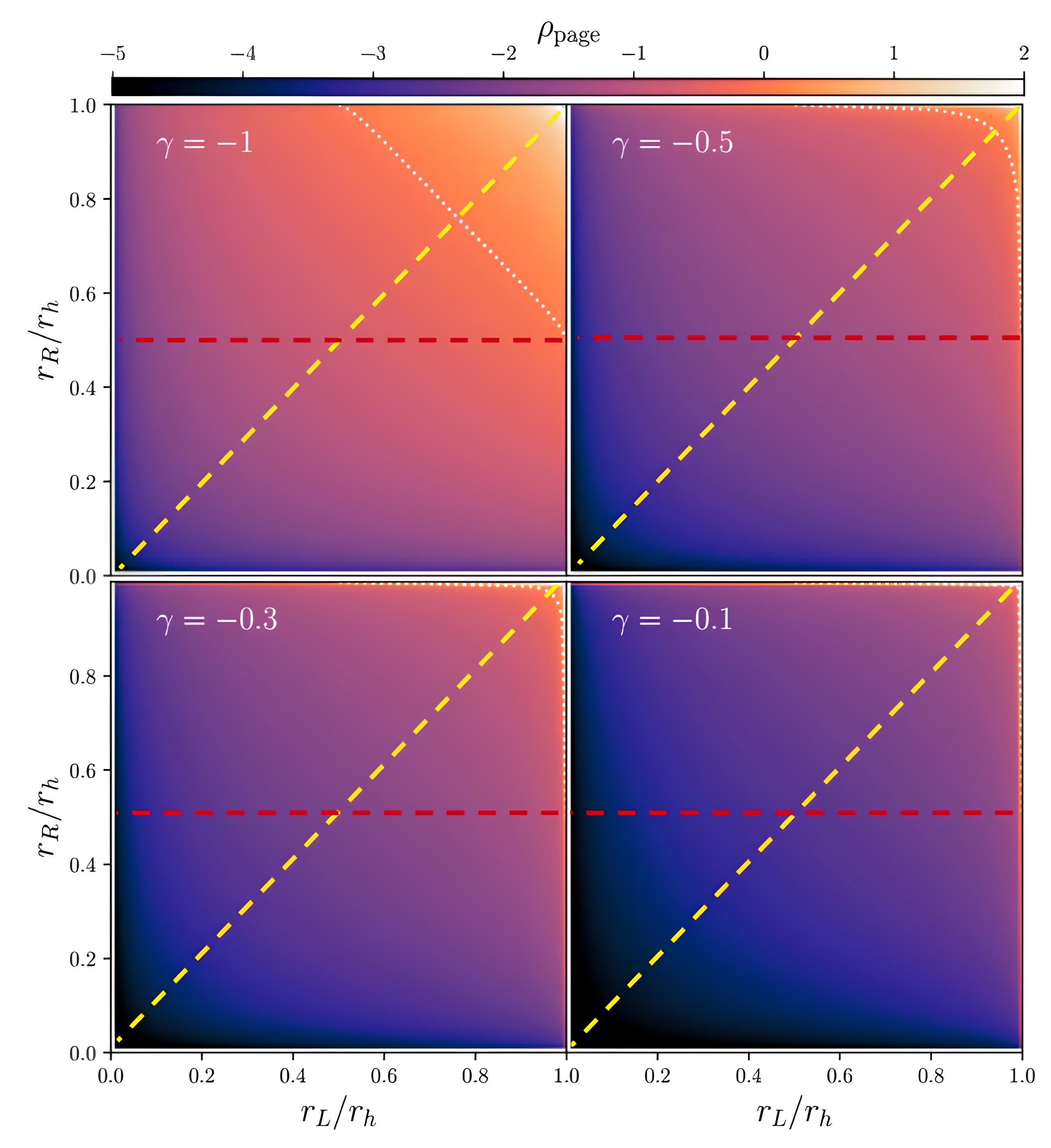}
\caption{The evolution of the density of the Page angle $\rho_{Page}$ (extracted and adapted from \cite{Santos:2024cvx}) around $r_L$ and $r_R$, for various values of $\gamma$ and fixed $\alpha=-8/3$. The contour with $\rho_{Page}=0$ is depicted by a white dotted curve as a reference.} \label{fig:rho_page_gamma}
\end{figure}
Through the conditions Page time ($S_{HM}(t_{Page})=S_{island}$) and Page angle ($S_{HM}(t=0)=S_{island}$) we will draw attention to the linear growth of the entanglement entropy \ref{LG}, this issue was not addressed in \cite{Santos:2024cvx} and here we draw attention to a behavior similar to that presented by \cite{Hartman:2013qma}, but corrected by the parameters of Horndeski gravity. For this, we need to establish the conditions:

$\bullet$ Early Times ($t < t_{\text{scr}}$) Fig \ref{fig:stream1}: Before any Hawking radiation escapes into the surrounding heat bath, the island region is trivial, and the entropy is dominated by a semiclassical contribution from the matter enclosed within the cutoff surface. This contribution remains invariant under time evolution if the black hole forms from the collapse of a pure state; the entropy at this stage is effectively zero.

$\bullet$ Intermediate Times ($t_{\text{scr}} < t < t_{\text{Page}}$) Fig \ref{fig:stream1}: As the black hole begins to evaporate, Hawking quanta escape into the heat bath, creating entanglement between the outgoing radiation and the interior modes. During this phase, the entropy grows due to the increasing number of mixed interior modes. A nontrivial island region begins to form near the event horizon, capturing the mixed interior modes and purifying the outgoing radiation. The semiclassical entropy starts to decrease as the black hole shrinks, and the generalized entropy of the nontrivial island closely tracks the shrinking area of the black hole horizon. At this stage, the entropy ($S(t)$) is dominated by the quantum mechanical contribution, ($S_{\text{HM}}[\bar{\mathcal{R}}(t)]$), where ($\bar{\mathcal{R}}(t)$) represents the radiation region.

$\bullet$ Late Times ($(t > t_{\text{Page}}$) Fig \ref{fig:stream1}: After the Page time ($t_{\text{Page}}$), the island region grows sufficiently large to purify most of the outgoing radiation. As a result, the semiclassical entropy becomes negligible. Simultaneously, the black hole continues to shrink, and the entropy is now dominated by the area term, $(S(t) \approx \frac{\mathcal{A}_{HM}(t)}{4G_N}$), where $\mathcal{A}_h(t)$ is the area of the black hole horizon and $G_N$ is Newton's constant.

The entropy initially follows the contribution from the trivial island but transitions to the contribution from the nontrivial island after the Page time Fig \ref{fig:t_page_gamma}. This transition is crucial for reproducing the Page curve Fig \ref{fig:rho_page_gamma}, which resolves the apparent conflict between the unitarity of quantum mechanics and the semiclassical description of black hole evaporation. When computing the entropy of Hawking radiation, it is essential to account for the contributions from the black hole interior, as encoded in the island region Fig \ref{fig:stream1}.

\subsection{Linear growth of entanglement entropy}\label{LG}
We present entropy in the boundary and the bulk channels to study how entanglement entropy grows linearly until it saturates the thermal
value. Considering two-point function $\langle\Phi_n(r_R,t)\,\Phi_n(r_L,t)\rangle$ of the twist operator fields 
$\Phi_n(r,t)$ inserted in the two bipartition points $r_L$ and $r_R$ \cite{Santos:2024cvx,Geng:2022dua}. The boundary and the bulk channels are given, respectively, as
\begin{eqnarray}
&&S_{\mathcal{A}}=min(S_{bulk},S_{bdry})=\frac{\mathcal{A}_{bulk}}{4G_N}+\frac{\mathcal{A}_{boundary}}{4G_N}\\
&&\mathcal{A}_{bulk}=6\chi\,log\,\left[\frac{2r_h\Delta_r}{r^2}\left(1+\omega(\rho_{
\epsilon})\sqrt{\Delta_L\,\Delta_R}\,\cosh\left(\frac{4\pi\,t}{\beta 
}\right)\right)\right]+2\chi\,log\left(\frac{2}{\epsilon}\right),\\
&&\mathcal{A}_{boundary}=8\,log(g_b)+2\chi\log\left(\frac{2}{\epsilon}\right),\\
\end{eqnarray}
where $\epsilon$ is the UV cutoff. for $t\gtrsim \beta$, the entanglement entropy grows linearly in time 
\begin{eqnarray}
S_{bulk}\to\chi\frac{t}{\beta}+S_{divergent};\,S_{divergent}=2\chi\,log\left(\frac{2}{\epsilon}\right).
\end{eqnarray}
Boundary entropy contributions according to \cite{Santos:2021orr} are similar to the AdS/BCFT black hole with corrections of Horndeski gravity:
\begin{eqnarray}
&&S^{L}_{bdry}=\frac{2}{3r_{L}}\left(1-\frac{\xi}{8}\right)+\frac{\xi}{3r^{2}_{L}}
\left(1-\frac{\xi}{4}\right)+\frac{\xi}{r^{2}_{L}}-\frac{\xi}{6}\\
&&S^{R}_{bdry}=\frac{2}{3r_{R}}\left(1-\frac{\xi}{8}\right)+\frac{\xi}{3r^{2}_{R}}
\left(1-\frac{\xi}{4}\right)+\frac{\xi}{r^{2}_{R}}-\frac{\xi}{6},
\end{eqnarray}
If we consider that the subsystems are large and far from the boundary, we can obtain
\begin{eqnarray}
&&S^{L}_{bdry}=-\frac{\xi}{6}\\
&&S^{R}_{bdry}=-\frac{\xi}{6}.
\end{eqnarray}
The entanglement entropy of a system initially grows linearly with time before saturating at its thermal equilibrium value

\begin{eqnarray}
&&S_{\mathcal{A}}=min(S_{bulk},S_{bdry})=S_{bulk}+S_{bdry}\\
&&S_{bulk}\to\chi\frac{t}{\beta}+S_{divergent};\,t\lesssim L,\\
&&S_{bdry}\to-\frac{\xi}{6}+S_{divergent};\,t \gtrsim L.
\end{eqnarray}
The transition between these two regimes typically occurs over a timescale of order $\beta$, where $\beta$ denotes the inverse temperature of the system. In contrast, for theories with gravitational duals, this transition is characterized by a sharp discontinuity, reflecting the underlying geometric structure of the dual spacetime. This sharp transition, provides valuable insights into the holographic correspondence and the dynamics of quantum information in strongly coupled systems \cite{Santos:2024cvx,Hartman:2013qma,Geng:2022dua}.

\section{Conclusion and discussions}\label{BHT2}

We probe that the thermal and entanglement entropy provide minimal information in terms of the entropy. The linear time dependence of entanglement entropy in black hole systems has profound implications for the evaporation process. As the entanglement entropy grows linearly with time, it suggests that the behavior of entanglement entropy at the initial time provides minimal information about the system $(S_{\mathcal{A}}\to-\frac{\xi}{6})$, the black hole does not evaporate completely, which is physical and consistent with our understanding of black hole thermodynamics ($S_{BH}\to-\frac{\xi}{6}$). This behavior likely arises from boundary Lagrangian ($\mathcal{L}_{boudnary}$), which corrects the entropy dynamics, as in realistic scenarios, to the entanglement entropy for end stages of evaporation \cite{Faulkner:2013ica,Maldacena:2015waa}. The scenario presented here captures the transition from a semiclassical description to a fully quantum mechanical one. Furthermore, the study of entanglement entropy in evaporating black holes is closely tied to the resolution of the black hole information paradox and the Page curve, which describes the entropy evolution of Hawking radiation \cite{Almheiri:2013hfa,Almheiri:2020cfm,Dong:2020uxp}.

Furthermore, the study of entanglement entropy in evaporating black holes is closely tied to the resolution of the black hole information paradox and the emergence of the Page curve. The Page curve, which describes the entropy evolution of Hawking radiation, is a critical feature in reconciling the principles of unitarity with the semiclassical description of black hole evaporation \cite{Almheiri:2013hfa, Almheiri:2020cfm, Dong:2020uxp}. Our analysis shows that the relationship between entanglement islands and Hawking radiation is essential for deriving the Page curve. The entanglement islands, which connect the black hole interior to the radiation region, ensure the conservation of information and provide a mechanism for studying the paradox.

Our findings also highlight the duality between smooth black hole interiors and firewall-like configurations. Only a subset of conformal field theory (CFT) microstates corresponds to black holes with smooth interiors, where the entanglement entropy evolves in a manner consistent with semiclassical expectations. In contrast, other microstates involve firewalls, where the semiclassical description breaks down. This duality underscores the richness of the CFT microstate structure and its implications for black hole physics. 

\acknowledgments
The author would like to thank Keryn Johnson, David Levitt, Paul Stowe, Hans Deyssenroth, Barry Mingst, Rulin Xiu, Jack Kiperman, Mei Yin, Dennis Zamudio Flores, and Sean Stotyn for the fruitful discussions during $3^{rd}$ International Summit on Gravitation, Astrophysics and Cosmology (ISGAC2025), 2025 in Valencia, Spain. 

\section{Data availability Statement}

The Data Availability Statement for this work isn't applicable because this manuscript is only being submitted for the first time at Fortschritte der Physik. 

\section{Conflict-of-Interest Statement}

The author declare that has no conflicts of interest related to this work.

\section{Statement of Funding}

This work is partially supported by Conselho Nacional de Desenvolvimento Cient\'{\i}fico e Tecnol\'{o}gico (CNPq) under grant 302835/2024-5.


\end{document}